\documentclass[aps,prd,preprintnumbers,twocolumn,groupedaddress,superscriptaddress,floatfix,10pt,longbibliography]{revtex4-1}
\usepackage{lipsum}
\usepackage{dcolumn}
\usepackage{bm}
\usepackage{xcolor}
\usepackage{amsfonts}
\usepackage{amsmath,amssymb}
\usepackage{graphicx}
\usepackage{booktabs}
\usepackage{orcidlink}
\usepackage{mathtools}
\usepackage{physics}
\usepackage{microtype}
\usepackage{siunitx}
\usepackage{multirow}

\numberwithin{equation}{section}

\usepackage{hyperref}
\hypersetup{colorlinks=true, linkcolor=blue, citecolor=blue, urlcolor=blue}

\begin{document}
\preprint{BARI-TH/791-26}

\title{Vector-meson properties in a data-driven AdS/QCD model}
\author{Xun Chen\,\orcidlink{0000-0001-8792-2398}}
\email{chenxun@usc.edu.cn}
\affiliation{School of Nuclear Science and Technology, University of South China, Hengyang 421001, China}
\affiliation{INFN --- Istituto Nazionale di Fisica Nucleare --- Sezione di Bari, Via Orabona 4, 70125, Bari, Italy}

\author{Floriana Giannuzzi\,\orcidlink{0000-0001-7482-7420}}
\email{floriana.giannuzzi@ba.infn.it}
\affiliation{INFN --- Istituto Nazionale di Fisica Nucleare --- Sezione di Bari, Via Orabona 4, 70125, Bari, Italy}

\author{Stefano Nicotri\,\orcidlink{0000-0003-2707-1098}}
\email{nicotri@infn.it}
\affiliation{INFN --- Istituto Nazionale di Fisica Nucleare --- Sezione di Bari, Via Orabona 4, 70125, Bari, Italy}
%\date{\today}

\begin{abstract}
We investigate the properties of $\rho, \psi,$ and $\Upsilon$ mesons within a holographic AdS/QCD framework, capturing both their vacuum phenomenology and in-medium dynamics.
By reconstructing flavor-specific gauge kinetic functions within a model originally developed to study thermodynamic functions, we compute the vector-meson mass spectrum, the decay constants and the hadronic vacuum polarization contributions to the anomalous magnetic moment of the muon at zero temperature. 
Furthermore, we evaluate the thermal spectral functions to analyze mass shifts, resonance broadening, and the melting of vector bound states in a hot and dense medium. 
Finally, using the Kubo formula, we extract the electrical conductivity of the plasma as a function of temperature and chemical potential.
\end{abstract}

\maketitle

\section{Introduction}
Exploring the low-energy regime of Quantum Chromodynamics (QCD) is still a profound theoretical challenge due to the onset of nonperturbative phenomena, such as color confinement and chiral symmetry breaking.
Over the past two decades, the Anti-de Sitter/Conformal Field Theory (AdS/CFT) correspondence \cite{Maldacena:1997re,Gubser:1998bc,Witten:1998qj} and its phenomenological extension (AdS/QCD) have emerged as powerful analytical tools. 
By mapping strongly coupled gauge theories in $d$ dimensions to weakly coupled gravity duals in $d+1$ dimensions, holographic models provide a highly tractable framework for computing hadronic observables.

Vector mesons play a central role in both strong interaction phenomenology and precision tests of the Standard Model (SM). 
In the AdS/QCD framework, the properties of these bound states are encoded in the two-point correlation function of the conserved vector current. 
One can extract fundamental vacuum properties, including the vector-meson mass spectrum and decay constants, by computing such a correlation function using holographic methods.
A precise theoretical determination of these quantities is of immediate phenomenological interest, particularly in the context of the Hadronic Vacuum Polarization (HVP) contribution to the anomalous magnetic moment of the muon, $a_\mu$. 
Because the HVP is entirely dominated by low-energy vector resonances (most notably the $\rho$ meson), holographic evaluations of the two-point function offer an independent methodology to constrain the theoretical uncertainties surrounding the muon $g-2$ puzzle.
In light of the limited quantitative predictivity typical of holographic models, it remains of great interest to explore whether introducing more robust phenomenological constraints into the meson sector can improve theoretical predictions for this quantity. 

Beyond the vacuum, understanding the fate of hadrons in extreme environments is important for assessing the features of QCD at finite temperature and density and the phase diagram. 
Ultra-relativistic heavy-ion collisions provide a valuable experimental laboratory to study nuclear matter under extreme conditions, leading to the formation of the Quark-Gluon Plasma (QGP). 
This deconfined state of quarks and gluons is not a weakly interacting gas, but rather a nearly perfect fluid. 
Indeed, experimental data concerning the ratio of the shear viscosity to entropy density ($\eta/s$) have demonstrated that the dynamics within the QGP is strongly coupled, so traditional perturbative QCD approaches are inherently limited for investigating this state of matter. 
Holographic approaches are suitable for this task instead, as finite temperature $T$ and baryon chemical potential $\mu$ can be straightforwardly incorporated by introducing a charged black hole in the bulk geometry.
In the thermal and dense medium obtained in this way, the vector two-point function is modified, allowing for the computation of the spectral function at finite temperature and density. 
The evolution of the spectral function serves as a diagnostic tool for in-medium effects, detailing how vector-meson bound states experience mass shifts, resonance broadening, and eventual melting into the deconfined plasma.
Furthermore, the spectral function bridges the gap between microscopic particle properties and the macroscopic transport coefficients of the QGP. 
According to the Kubo formula, the slope of the vector spectral function at vanishing frequency and momentum directly dictates the electrical conductivity of the strongly interacting medium. 
Evaluating the conductivity at finite $T$ and $\mu$ helps to understand the generation and relaxation of electromagnetic fields in heavy-ion collisions, and holographic AdS/QCD duality provides an effective framework to compute such transport coefficients in the strongly coupled regime  \cite{Cao:2021tcr, MartinContreras:2021bis,Zhao:2021ogc,MartinContreras:2026uyt,Finazzo:2015xwa,Finazzo:2013efa}.

In this paper, we present a comprehensive investigation of vector meson properties within a data-driven bottom-up AdS/QCD model. 
In particular, we adopt the model of Ref.~\cite{Chen:2025goz}, and extend it in order to include the fluctuations of the vector meson fields describing $\rho$, $\psi$, and $\Upsilon$ mesons in the vacuum and at finite temperature and chemical potential.
To achieve this, we define one gauge kinetic function for each flavor-current sector, shared by the full radial tower, encoding the coupling of vector mesons with a dilaton field. 
They are reconstructed using a smooth-basis ansatz and the following observables as training input: the masses and decay constants of the first two resonances, and the value of the gluon condensate extracted from the operator product expansion (OPE) of the two-point function.
We compute the mass, decay constant, and HVP contribution in the zero-temperature limit, and subsequently extend the analysis to a hot and dense background to evaluate the in-medium spectral function and electrical conductivity.

The paper is organized as follows. In Section II, we introduce the holographic setup by fixing the metric of the $5d$ spacetime in which the vector field is defined, and we employ a gauge kinetic function fitted using thermodynamic observables to investigate $\rho$ meson properties. 
In Section III, we present three new, purpose-built gauge kinetic functions to analyze the properties of light mesons, charmonia, and bottomonia. Finally, Section IV is dedicated to our conclusions and outlook.

\section{Review of the holographic model}
We investigate vector-meson phenomenology in the AdS/QCD model introduced in Ref.~\cite{Chen:2025goz} to study the equation of state and the QCD phase diagram.
It is an analytical five-dimensional bottom-up holographic model of QCD whose parameters are refined fitting to $(2+1)$-flavor lattice data via gradient descent optimization to precisely locate the critical endpoint at finite temperature and density. 
It consists of an Einstein-Maxwell-Dilaton (EMD) framework with a $5d$ AdS-like black-hole background metric, which in the Einstein frame is generically taken as
\begin{equation}\label{eq:metric}
\mathrm{d}s^2=\frac{L^2 e^{2A(z)}}{z^2}\left[-g(z)\,\mathrm{d}t^2+\frac{\mathrm{d}z^2}{g(z)}+\mathrm{d}\vec{x}^{\,2}\right],
\end{equation}
where $L$ is the AdS radius, $z$ is the ``holographic'' coordinate ($z\to 0$ corresponds to the UV boundary), and $g(z)$ is the blackening function.
The black-hole horizon is located at $z=z_h$ satisfying $g(z_h)=0$. 
In this model, temperature is defined as
\begin{equation}
    T = \displaystyle\frac{\abs{g'(z_h)}}{4\pi}\,\,.
\end{equation}
Conformal symmetry is broken by introducing a scalar (dilaton) field $\phi(z)$, which effectively produces a nontrivial infrared scale, linked to the QCD scale $\Lambda_{\mathrm{QCD}}$, leading to a deformed geometry for $z>0$, described by a nontrivial warp factor $A(z)$.

The massless quark limit of QCD is characterized by a global chiral symmetry, $U(n_f)_L \times U(n_f)_R$, where $n_f$ is the number of quark flavors. 
According to the holographic dictionary, a global symmetry on the $4d$ boundary corresponds to a local gauge symmetry in the $5d$ bulk. 
Therefore, to holographically describe chiral symmetry, the model involves two $5d$ gauge fields in the bulk: $L_M$ (for the left-handed chiral sector) and $R_M$ (for the right-handed chiral sector).
These left/right fields can then be rotated into vector/axial-vector ones: $L_M = V_M + A_M $ and $R_M = V_M - A_M $. 
The vector field is dual to the conserved QCD vector current $\bar q \gamma^\mu q$, whose excitations represent the vector meson nonet.
The chiral symmetry group $U(n_f)_L \times U(n_f)_R$ can be reorganized into vector and axial-vector transformations. The vector sector forms the subgroup $SU(n_f)_V \times U(1)_V$, where $U(1)_V$ ensures baryon number conservation, while the axial transformations are generated by $SU(n_f)_A$ and $U(1)_A$, the latter being explicitly broken by the axial anomaly.
In matrix representation, $V_M=V_M^A T^A$, with $A=0,1,...8$ for $n_f=3$, and $T^A$ the generators of $U(3)_V$, such that $\mathrm{Tr}[T^AT^B]=\delta^{AB}/2$.

The action considered in \cite{Chen:2025goz}, involving the gravitational field $g_{MN}$, and the $U(1)_V$ and dilaton fields, is given by
\begin{equation}
    \begin{aligned}
        S = \frac{1}{16 \pi G_5} \int \mathrm{d}^5 x \sqrt{-g} & \left[ R - \frac{f(\phi)}{4} F^2 \right. \\
        & \left. - \frac{1}{2} \partial_M \phi \partial^M \phi - \mathcal{V}(\phi) \right],
    \end{aligned}
\end{equation}
where $f(\phi)$ is the gauge kinetic function which controls the coupling between the dilaton and gauge fields, and $\mathcal{V}(\phi)$ is the dilaton potential.
The action of the full $U(3)_V$ field is:
\begin{eqnarray}\label{eq:SVthermo}
S_V
&=&-\frac{1}{16 \pi G_5}\int \mathrm{d}^5x\,\sqrt{-g}\,\frac{f(\phi)}{2}\,\mathrm{Tr}\!\left[F_V^{MN}F^V_{MN}\right]\,. \nonumber\\
\end{eqnarray}
We neglect terms proportional to the scalar field $X(z)$, representing the vacuum expectation value (vev) of the scalar operator $\bar qq$, which contains the chiral condensates and quark masses, that contribute to the  action of the vector field $V_M^b$ with $b=4,5,6,7$ when chiral symmetry is explicitly broken.
The action $S_V$ contains the time component of the vev of the $U(1)_V$ gauge field, which backreacts on the metric.
%$B_0(z) \frac{1}{3} \mathbb{I}_{3\times3}$, where $\mathbb{I}_{3\times3}$ is the identity matrix
The holographic source of this field is the baryon chemical potential $\mu_B$, related to the quark chemical potential $\mu_q$ by $\mu = \mu_B = 3\mu_q$.
The dynamical fluctuations $V^A(x,z)$ of the vector field are introduced in the probe approximation, \emph{i.e.} without considering their backreaction on the metric. 

Following~\cite{Chen:2025goz}, the warp factor is given by:
\begin{equation}\label{eq:Az}
    A(z)=d\,\log(1+a z^2)+d\,\log(1+b z^4)\,,
\end{equation}
with $a=0.2219~\mathrm{GeV}^2$, $b=0.033~\mathrm{GeV}^4$, and $d=-0.1212$. 
This choice ensures the asymptotic AdS behavior in the UV and introduces a controlled deformation in the IR.
The gauge kinetic function, parametrized as a function of $z$, is:
\begin{equation}\label{fff}
    f(z)=\frac{3}{2}\frac{k\,  e^{c z^2+ h z^4 + n z^6 -A(z)}}{-2c- 4 h z^2-6 n z^4}\,,
\end{equation}
with $k= 0.02243$ GeV$^{2}$, $c= -0.0358$ GeV$^{2}$, $h= 0.0039$ GeV$^{4}$, and $n= -0.0010$ GeV$^{6}$.
We have rescaled the gauge-kinetic function $f(z)$ by a factor $3/2$ relative to Ref. \cite{Chen:2025goz} to account for the different normalization of the baryon gauge field. While Ref. \cite{Chen:2025goz} employs an ordinary Abelian field with $A_0(0)=\mu_B$, here the baryon background is embedded in $U(3)_V$ as $V_0(0)=\frac{\mu_B}{3} \mathbb{I}_{3\times3}$ \cite{Bartolini:2023wis}.

The last parameter appearing in the action is $G_5/L^3=0.4113$. 
In the following $L=1$ will be used.
The parameters have been determined in \cite{Chen:2025goz} using an iterative process involving comparison to lattice results, with a gradient descent optimizer used to automatically adjust them in order to minimize the difference between holographic outcomes and lattice calculations. 
The observables used to calibrate the parameters of the model are the speed of sound at $\mu_B = 0$, 
the second-order baryon number susceptibility, 
and the baryon number density at $\mu_B/T = 1$.

\subsection{Vector mesons}\label{sec:therm_mesons}
Let us study the $\rho$ spectrum at zero temperature and chemical potential, so we use $g(z)=1$.
It is convenient to write $\mathrm{d}s^2 = e^{ 2A_e(z)}(\eta_{\mu\nu}\mathrm{d}x^\mu \mathrm{d}x^\nu + \mathrm{d}z^2)$ with $e^{2A_e}\equiv L^2 e^{2A}/z^2$. 
In axial gauge $V_z=0$ and transverse polarization, the mode equation obtained from the Lagrangian in \eqref{eq:SVthermo} reads
\begin{equation}\label{eq:vectoreom}
\partial_z\!\left(e^{A_e(z)} f(z)\, \partial_z \psi(z)\right) + m^2\, e^{A_e(z)} f(z)\, \psi(z)=0\,,
\end{equation}
where $\psi(z)$ denotes one component of the vector field, and $m^2=-q^2$ is the meson mass. 

Define the positive weight function
\begin{equation}
\mathcal{P}(z)\equiv e^{A_e(z)} f(z), 
\qquad 
\hat{\psi}(z)\equiv \sqrt{\mathcal{P}(z)}\,\psi(z),
\end{equation}
so that Eq.~\eqref{eq:vectoreom} can be recast into the Schr\"odinger form
\begin{equation}\label{eq:Sch}
-\hat{\psi}''(z) + V_V(z)\,\hat{\psi}(z) = m^2\, \hat{\psi}(z),
\end{equation}
with the effective potential
\begin{equation}\label{eq:VV}
V_V(z)=\frac{1}{2}\frac{\mathcal{P}''(z)}{\mathcal{P}(z)}
-\frac{1}{4}\left(\frac{\mathcal{P}'(z)}{\mathcal{P}(z)}\right)^2.
\end{equation}

The decay constant $f_{\rho_n}$ of the $n$-th neutral $\rho$ meson state, denoted as $\rho_n$, is defined through the matrix element of the vector current $J^3_\mu = \bar q T^3 \gamma_\mu q$ between the hadronic vacuum and the meson state:
\begin{equation}\label{eq:definitionDecCnst}
    \langle 0 | \frac{1}{2} (\bar{u}\gamma_\mu u - \bar{d}\gamma_\mu d) | \rho_n^0(p, \lambda) \rangle = m_{\rho_n} f_{\rho_n} \epsilon_\mu(p, \lambda) \,,
\end{equation}
where $m_{\rho_n}$ is the mass of the specific $\rho$ resonance, and $\epsilon_\mu(p, \lambda)$ is the polarization vector of the meson with momentum $p$ and helicity $\lambda$.
In AdS/QCD models, decay constants can be computed from \cite{Erlich:2005qh}:
\begin{equation}\label{eq:DecCnst}
    f_{\rho_n} = \lim_{z\to 0} \frac{1}{(16\pi G_5)^{1/2} \, m_{\rho_n}} \mathcal{P}(z) \psi_n'(z) \,.
\end{equation}

The values obtained for eigenvalues $m_{\rho_n}$ and decay constants are indicated in Table \ref{tab:rho_thermo}. 
Masses are higher than experimental data, and the decay constant of the ground state is smaller than the expected value $f_{\rho_1}=155$ MeV.
These discrepancies suggest that the  gauge kinetic function, tuned to successfully reproduce QCD thermodynamics from lattice data, cannot simultaneously describe the vector meson spectrum with the same level of accuracy.
In the next section we shall determine which behavior the function $f(\phi)$ should have in order to reproduce vector-meson spectrum, emphasizing differences with \eqref{fff}.

\begin{table*}[tbp]
\centering
\caption{Masses and decay constants (in MeV) of $\rho$-meson $n S$ radial excitations found in the EMD model of Section \ref{sec:therm_mesons} (Th mass, Decay constant) and from PDG (Exp mass) \cite{ParticleDataGroup:2026aaa}.}
\label{tab:rho_thermo}
\begin{ruledtabular}
\begin{tabular}{c c c c c}  
\textbf{State} & \textbf{Meson} & \textbf{Exp mass} & \textbf{Th mass} & \textbf{Decay constant} \\
\colrule
\rule{0pt}{3ex}
$1S$ & $\rho(770)$ & $775.26 \pm 0.23$ & 1092 & 131\\
$2S$ & $\rho(1450)$ & $1465 \pm 25$ &  1885 & 121 \\
$3S$ & $\rho(1900)$ & $1880 \pm 10$ (Ablikim 2022) & 2602 & 147 \\
%\multirow{2}{*}{$3S$} 
%& $\rho(1700)$ & $1720 \pm 20$   & --- & --- \\
%& $\rho(1900)$ & $1880 \pm 10$ (Ablikim 2022) & 2602 & 147 \\
\end{tabular}
\end{ruledtabular}
\end{table*}

From Eq.~\eqref{eq:SVthermo}, the two-point function of vector mesons can be computed by deriving twice the on-shell action with respect to the source of the vector field.
In the Fourier space, the field $V_\mu$ is related to its source $V_{0\mu}$ through
\begin{equation}
    V_\mu(q_0^2,\bar q^2,z) = V(z,q_0^2,\bar q^2) V_{0\mu}(q_0^2,\bar q^2)  \,,
\end{equation}
where $V(z,q_0^2,\bar q^2)$ is the bulk-to-boundary propagator, that is computed by solving Eq.~\eqref{eq:vectoreom} with $m^2=-q^2$ and boundary conditions $V(0,q_0^2,\bar q^2)=1$ plus regularity for $z\to\infty$.
The two-point correlation function is given by:
\begin{equation}
    \Pi_{\mu\nu}(q^2) = \frac{\partial^2 S_{os}}{\partial V_0^\mu \partial V_0^\nu} = (q_\mu q_\nu - q^2 \eta_{\mu\nu}) \Pi(q^2)
\end{equation}
with
\begin{equation}\label{eq:twoptfunc}
    \Pi(q^2) = -\frac{1}{16 \pi\ G_5} \lim_{z\to 0} \frac{e^{A(z)}}{q^2 z}\, f(z)\, V(z,q^2)\, \partial_z V(z,q^2)\,. 
\end{equation}
In the high $Q^2=q^2$ limit, it behaves as
\begin{equation}\label{eq:Pilogfthermo}
    \Pi(Q^2) \to -c_Q \log Q^2
\end{equation}
with $c_Q\sim 0.011$, slightly lower than the expected perturbative coefficient $N_c/(24\pi^2)$ \cite{Erlich:2005qh}.

Let us investigate how vector mesons behave in a thermal background.
In holographic models, finite temperature and density effects are incorporated by introducing a black hole in the $5d$ space. 
It is described by the blackening function $g(z)$ in Eq.~\eqref{eq:metric}, and the black-hole horizon $z_h$ is the value of the fifth coordinate such that $g(z_h)=0$. 

The variation of meson masses and widths with temperature and chemical potential can be extracted from the hadronic spectral function, defined as twice the imaginary part of the retarded Green's function \cite{Aarts:2020dda,Bellantuono:2014lra,Colangelo:2012jy,Colangelo:2009ra}:
\begin{equation}\label{eq:SF}
    \rho(q_0^2,\bar q^2) = 2 \Im \Pi_R (q_0^2,\bar q^2) \,.
\end{equation}
%In AdS/QCD, the Green's function is found by deriving twice the on-shell action with respect to the source of the field $V_\mu$. 
At finite temperature and density the equation of motion for the bulk-to-boundary propagator $V(z,q_0^2,\bar q^2)$ of the spatial component of the vector field reads:
\begin{equation}
    \partial_u\left( \frac{e^A}{u} \, f \, g\, \partial_u V\right) + \left( \frac{q_0^2}{g}-\bar q^2\right) \frac{e^A\, z_h^2}{u} \, f\, V=0 \,,
\end{equation}
where $u=z/z_h$.
Analogously to Eq.~\eqref{eq:twoptfunc}, the retarded Green's function at finite temperature and density can be computed from the on-shell action obtaining:
\begin{equation}\label{eq:greenfun}
    \Pi_R(q^2) = - \frac{1}{16 \pi\ G_5\, z_h^2} \lim_{u\to 0} \frac{e^A}{u}\, f\, g\, V\, \partial_u V\,, 
\end{equation}
requiring  the bulk-to-boundary propagator to behave as an in-falling wave near the black-hole horizon:
\begin{equation}
V(u,q_0^2,\bar q^2) \xrightarrow{u\to 1} (1-u)^{i q_0/g_1} (1+(1-u)...)   
\end{equation}
with $g_1=g'(z_h)$.

\begin{figure}[h!]
    \centering
    \includegraphics[width=0.8\linewidth]{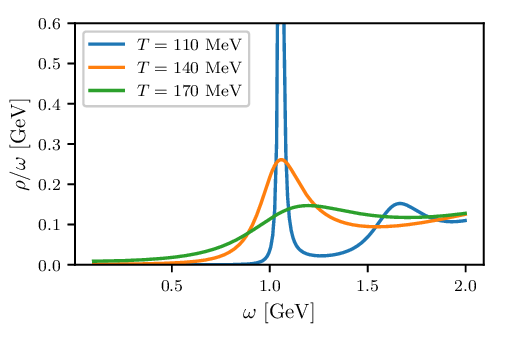}
    \caption{\label{fig:rhoSFmu0thermo} Spectral function of the $\rho$ meson for $\bar q=0$,  $\omega=q_0$, $\mu=0$ and $T=110$ MeV (blue curve), $T=140$ MeV (orange curve), $T=170$ MeV (green curve).
    The model is characterized by the gauge kinetic function in Eq.~\eqref{fff}.}
\end{figure}

\begin{figure}[h!]
    \centering
    \includegraphics[width=0.8\linewidth]{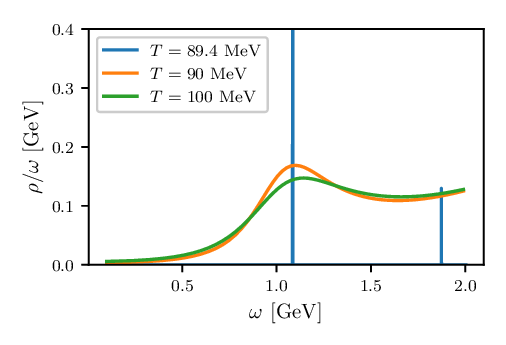}
    \caption{\label{fig:rhoSFmu1thermo} Spectral function of the $\rho$ meson for $\bar q=0$, $\mu=1$ GeV and $T=89.4$ MeV (blue curve), $T=90$ MeV (orange curve), $T=100$ MeV (green curve). The model is characterized by the gauge kinetic function in Eq.~\eqref{fff}.}
\end{figure}

The $\rho$ spectral function at  $\omega=q_0$ and $\bar q=0$ is shown in Figs. \ref{fig:rhoSFmu0thermo}-\ref{fig:rhoSFmu1thermo}  for $\mu=0$ and $\mu=1$ GeV, respectively, and for some values of temperature. 
The typical broadening of the peaks can be observed. 
In \cite{Chen:2025goz} it was found that the transition from a confined to a deconfined phase  at $\mu=0$ is a crossover, and it occurs in the range of temperatures 140-170 MeV.
Indeed, we find that the broadening of peaks in the spectral function at $\mu=0$ is continuous as well. 
At $\mu=0$ and $T=170$ MeV the ground state peak has  disappeared from the spectral function. 
Conversely, at $\mu=1$ GeV a first-order phase transition occurs at $T_c = 89.4$ MeV \cite{Chen:2025goz}. 
Fig.~\ref{fig:rhoSFmu1thermo} shows that the first resonance peak is narrow at $T=89.4$ MeV (immediately below the phase transition), while it suddenly becomes large at $T=90$ MeV.
Both examples show that the $\rho$ meson melts at temperatures slightly above the deconfinement phase transition.


%\subsection{Conductivity}
The electromagnetic spectral function $\rho_{ii}$, specifically its zero-frequency limit at vanishing three-momentum $\bar q$, can be used to determine the electrical conductivity $\sigma$ of the medium via the Kubo formula \cite{Aarts:2020dda}:
\begin{equation}
    \sigma= \frac{1}{6} \lim_{q_0\to 0} \frac{\sum_{i=1}^3\rho_{ii}(q_0,\bar q=0)}{q_0} \,,
\end{equation}
where the summation is over spatial components, $i = 1, 2, 3$, and $\rho_{ii}$ contains the factor $C_{\mathrm{em}}=e^2 \sum_f q_f^2$ representing the sum over $u,d,s$ electric charges.
The field dual to the electromagnetic current $J_\mu=e Q_{\mathrm{em}} \bar q \gamma_\mu q$, where $Q_{\mathrm{em}}=\mbox{ diag}(q_u,q_d,q_s)$ is the quark charge matrix, is again a vector field the source of which is proportional to the matrix $Q_{\mathrm{em}}$ \cite{Colangelo:2011xk,Colangelo:2023een}.
The three spatial components of this vector field are identical, so their spectral functions are equal as well: $\rho_{11}=\rho_{22}=\rho_{33}=\rho$, where $\rho$ is defined in Eq.~\eqref{eq:SF}.

The retarded Green's function of the spatial component of the electromagnetic current is given by \eqref{eq:greenfun}, with a different overall factor related to the trace of the source matrices, so in this case it reads:
\begin{equation}
    \Pi^R_{ii} = - \frac{1}{8 \pi G_5 z_h^2} \mbox{Tr}[e^2 Q_{\mathrm{em}}^2]  \lim_{u\to 0} \frac{e^A}{u} f\, g\, V \partial_u V \,.
\end{equation}
%Then, the spectral function $\rho_{ii}$ is found from Eq.~\eqref{eq:SF}.
Notice that $C_{\mathrm{em}}=\Tr [e^2 Q_{\mathrm{em}}^2]$.

Iqbal and Liu have shown that the low frequency limit of a strongly coupled field theory at finite temperature is determined by the horizon geometry of its gravity dual, and, in particular, boundary theory transport coefficients can be expressed in terms of geometric quantities evaluated at the horizon \cite{Iqbal:2008by}.
In this model, the analytical formula for the conductivity reads \cite{Iqbal:2008by}:
\begin{equation}\label{eq:sigma}
    \sigma = \frac{1}{8 \pi G_5}  \mbox{ Tr}[e^2 Q_{\mathrm{em}}^2] \frac{e^{A(z_h)} \, f(z_h)}{z_h}\,.
\end{equation}
As a test of the numerical computation of the spectral functions, we have checked that the extracted conductivity matches the analytical formula \eqref{eq:sigma}.

\begin{figure}[h!]
    \centering
    \includegraphics[width=0.8\linewidth]{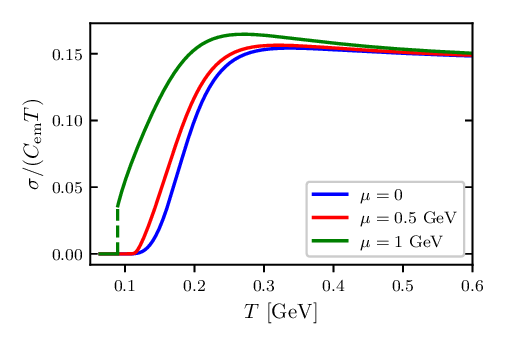}
    \caption{\label{fig:sigmatherm} Electrical conductivity $\sigma/(C_{\mathrm{em}} T)$ for $\mu=0$ (blue curve), $\mu=0.5$ GeV (red curve), $\mu=1$ GeV (green curve). 
    The model is characterized by the gauge kinetic function in Eq.~\eqref{fff}.}
\end{figure}

Fig.~\ref{fig:sigmatherm} shows the dimensionless ratio $\sigma/(C_{\mathrm{em}} T)$ as a function of temperature for $\mu=0$, $\mu=0.5$ GeV, $\mu=1$ GeV. 
In the latter case, the first-order phase transition at $T_C$ causes a discontinuity of the conductivity. 
At large temperatures, the conductivity is independent of the chemical potential, and it reaches a constant value $\sigma/(C_{\mathrm{em}} T)\sim 0.143$.

Lattice-QCD calculations with $N_f=2+1$ dynamical quark flavors indicate that $\sigma/(C_{\mathrm{em}} T)$ is suppressed in the vicinity of the deconfinement crossover and increases smoothly with temperature in the deconfined phase, reaching values of approximately $0.31 - 0.35$ at $T\sim 350$ MeV \cite{Amato:2013naa,Aarts:2014nba}.
These estimates are extracted by reconstructing the low-frequency vector spectral function from Euclidean lattice correlators, and are affected by sizeable systematic uncertainties.
A qualitatively similar temperature dependence is observed in kinetic quasi-particle approaches, in which medium effects are encoded through temperature-dependent effective masses, couplings, and relaxation times \cite{Puglisi:2014pda,Mykhaylova:2020pfk}.
The conductivity computed in our holographic model exhibits a significantly steeper rise at the deconfinement transition compared to the broader crossover characteristic of the lattice results. 
The finiteness of $\sigma/(C_{\mathrm{em}} T)$ in the high temperature regime provides strong evidence that the QGP behaves as a strongly coupled fluid. This contrasts  with the divergent behavior expected from weakly coupled perturbative QCD, but aligns with holographic calculations in strongly coupled $\mathcal{N}=4$ Supersymmetric Yang-Mills (SYM) theory, which yield a temperature-independent constant of $\sigma/T = N_c^2/16\pi$ \cite{Policastro:2002se,Caron-Huot:2006pee}.

\section{New gauge kinetic function}\label{sec:newgaugefunc}
In the previous section we have used the model developed in \cite{Chen:2025goz} to study vector-meson spectroscopy. 
We have found that, although the model provides a very good description of thermodynamic functions, it fails in providing an accurate determination of vector-meson masses, both for the ground state and radial excitations.
This could be due to the fact that the model parameters were tuned exclusively using finite temperature observables, and the description of vacuum properties may not be captured by this calibration.
We then wonder how the gauge kinetic function $f(\phi)$ should be modified to accurately reproduce the vector meson spectrum.
As argued in Ref.~\cite{He:2013qq}, the coupling function $f_V(\phi)$ in the probe-vector action is not strictly required to coincide with the gauge kinetic function $f(\phi)$ of the background EMD action. 
Consequently, we can treat $f_V$ as an independent function to be constrained directly by hadron spectroscopy.
Thus, we introduce a distinct function $f_V(\phi)$ coupled to the fluctuations of the vector field.
This approach parallels the original Karch-Katz-Son-Stephanov (KKSS) soft-wall model~\cite{Karch:2006pv}, where the vector sector couples to the background dilaton via an overall factor $e^{-\phi(z)}$:
\begin{equation}\label{eq:KKSS}
S_{\mathrm{KKSS}}
=-\frac{1}{2g_5^2}\int \mathrm{d}^5x\,\sqrt{-g}\,e^{-\phi(z)}\,\mathrm{Tr}\!\left[F_V^{MN}F^V_{MN}\right],
\end{equation}
where $F^V_{MN}=\partial_M V_N-\partial_N V_M - i [V_M,V_N]$ is the field strength associated with the (flavor) vector gauge field $V_M(x,z)$.
In the soft-wall model the metric was pure AdS, and the dilaton was a non-dynamical field.
We generalize this coupling by introducing a profile $f_V(\phi)$,
\begin{equation}\label{eq:SV}
S_\rho
=-\frac{1}{2g_5^2}\int \mathrm{d}^5x\,\sqrt{-g}\,f_V(\phi)\,\mathrm{Tr}\!\left[F_V^{MN}F^V_{MN}\right],
\end{equation}
with $f_V=1$ on the boundary and $g_5^2=12\pi^2/N_c$ \cite{Erlich:2005qh}.
Furthermore, this procedure allows for the description of both light and heavy mesons, specifically charmonia and bottomonia, simply by identifying a distinct, flavor-dependent gauge kinetic function for each sector.

To make contact with the soft-wall notation, it is convenient to introduce an \emph{effective vector dilaton} profile $\phi_V(z)$ via
\begin{equation}\label{eq:fV_def}
f_V(\phi(z)) \equiv e^{-\phi_V(z)}.
\end{equation}
For brevity, we shall often write $f_V(z)$ instead of $f_V(\phi(z))$.
We parametrize the dilaton  as
\begin{align}\label{eq:phiV_def}
  \phi_V(z)&=\kappa\, z^2+\sum_{i=1}^2 \sum_\pm a_i\left[\mathrm{Sp}\left(\frac{\pm z-c_i}{s_i}\right)-\mathrm{Sp}\left(\frac{-c_i}{s_i}\right)\right] \nonumber \\
       &
\end{align}
with $\mathrm{Sp}(x) = \log(1+e^x)$ the softplus function.
The function satisfies $\phi_V(0)=0$, $f_V(0)=1$, and ensures that the gauge kinetic function contains only even powers of $z$ as $z\to0$.
This condition is particularly important, since odd terms would spoil holographic renormalizability of the action and conformality of the boundary CFT, and since it guarantees a correct $1/Q^2$ expansion of the two-point function.
We have chosen to sum only two softplus functions since we expect that the kinetic function has a simple shape. 
This prevents overfitting and makes the model more predictive.

In this data-driven setup, $\phi_V(z)$ will be reconstructed directly from the vector-meson spectrum and decay constants, in the background described by the metric in Eqs. \eqref{eq:metric} and \eqref{eq:Az}. 
Masses constrain the behavior of the function at large $z$, while decay constants allow us to control the behavior at small $z$.
To get less ambiguity in reconstructing the function, we also evaluate the gluon condensate, imposing that it must approach the value $\langle \frac{\alpha_s}{\pi}G^2\rangle \sim (0.012\pm 0.0036)$ GeV$^4$ \cite{Colangelo:2000dp}.
Masses and decay constants of vector mesons are computed, respectively, as the eigenvalues of Eq.~\eqref{eq:Sch} and from Eq.~\eqref{eq:DecCnst}, replacing $f(z)$ with the new gauge function $f_V(z)$ and $16\pi G_5$ with $g_5^2$.

We extract the value of the gluon condensate from the Operator Product Expansion (OPE)  of the two-point function \eqref{eq:twoptfunc}, which in this case reads:
\begin{equation}\label{eq:twoptfuncNEW}
    \Pi(Q^2) = -\frac{1}{g_5^2 Q^2} \lim_{z\to 0} \frac{e^A}{z}\, f_V\, V\, \partial_z V\,,
\end{equation}
where $V(z,Q^2)$ is the bulk-to-boundary propagator.
In order to get an analytical expression of the OPE, we adopt the Green's function method described in Ref.~\cite{Colangelo:2011xk}. 
We seek an asymptotic $Q^2\to\infty$ solution to the equation of motion \eqref{eq:vectoreom} with $m^2 = -Q^2$. 
To achieve this, we introduce the rescaled variable $t = z \sqrt{Q^2}$, such that the equation becomes:
 \begin{multline}\label{eq:eomint}
     \partial_t^2 V(t,Q^2) -\frac{1}{t} \partial_t V(t,Q^2)-V(t,Q^2) =\\
     -\partial_t (A-\phi_V) \partial_t V(t,Q^2)\,.
\end{multline}
The first factor on the right-hand side can be expanded in $1/Q^2$ as:
\begin{equation}
    A -\phi_V \sim \frac{a_{f2}}{Q^2} t^2 + \frac{a_{f4}}{Q^4} t^4 + \mathcal{O}(t^6/Q^6)\,,
\end{equation}
and the general solution as
 \begin{equation}
     V(t,Q^2) = \sum_{n} \left(\frac{1}{Q^2}\right)^n V_n(t)\,.
 \end{equation}
 By solving Eq.~\eqref{eq:eomint} order by order we find:
 \begin{align}
     V_0(t) =& t K_1(t) \nonumber\\
     V_1(t) =& -\frac{t^3}{2} a_{f2} K_1(t) \nonumber\\
     V_2(t) =& \frac{t^3}{24} \big(-4 (a_{f2}^2 + 4 a_{f4}) t K_0(t) \nonumber\\
     & +( a_{f2}^2 (-8+3t^2 )-4 a_{f4} (8+3t^2)) K_1(t)\big) \nonumber\\
 \end{align}
where $K_i(t)$ is the modified Bessel function of the second kind.
Then, the high-$Q^2$ expansion of the two-point function of the vector current $\bar q T^a \gamma^\mu q$ is obtained:
\begin{align}\label{eq:PihighQAdS}
    \Pi(Q^2) \xrightarrow[Q^2\to\infty]{} & \frac{1}{4\pi^2} (\log(2\nu)-\gamma_E) - \frac{1}{8\pi^2} \log Q^2 \nonumber\\
    &+\frac{1}{4\pi^2} \frac{a_{f2}}{Q^2} +\frac{1}{6\pi^2} \frac{a_{f2}^2+4 a_{f4}}{Q^4}+ \mathcal{O}(Q^{-6})\,,
\end{align}
with
\begin{equation}
    a_{f2}=a\, d - \kappa - \sum_{i=1}^2 \frac{a_i \, \mathrm{sech}^2(c_i/(2 s_i))}{4\, s_i^2}
\end{equation}
 \begin{multline}
    a_{f4} = -\frac{a^2\, d}{2} + b\, d \\
    - \sum_{i=1}^2 \frac{a_i\, (-2 + \mathrm{cosh}(c_i/s_i)) \mathrm{sech}^4(c_i/(2 s_i))}{96\, s_i^4}\,.
\end{multline}
The same expression holds for the vector current of heavy quarks $\bar c \gamma^\mu c$ and $\bar b\gamma^\mu b$, multiplied by a factor 2, due to the absence of the trace of Gell-Mann matrices.  

Let us compare the holographic expression of the OPE of the two-point function to the one found from QCD sum rules, which in the light-quark sector reads  \cite{Shifman:1978bx}:
\begin{eqnarray}\label{eq:PirhohighQQCD}
    \Pi^{u,d,s}_{\mbox{QCD}}(Q^2) &\sim& -\frac{1}{8\pi^2} \left( 1+\frac{\alpha_s}{\pi}\right) \log \frac{Q^2}{\nu^2} -\frac{3 m_q^2}{4\pi^2 Q^2} \nonumber\\
    &&+\frac{1}{Q^4} \left(\langle m_q \bar q q\rangle + \frac{1}{24} \langle \frac{\alpha_s}{\pi} G^2\rangle \right)+ ... \nonumber\\
\end{eqnarray}
$\nu$ being a renormalization scale.
In the heavy-quark sector there is no underlying chiral symmetry, and the heavy-quark condensate reduces to the gluon condensate \cite{Dominguez:2014pga}:
\begin{equation}
    \langle \bar Q Q \rangle = -\frac{1}{12 m_Q} \langle \frac{\alpha_s}{\pi} G^2\rangle\,.
\end{equation}
Therefore the OPE of the two-point function of vector currents containing heavy quarks is \cite{Shifman:1978bx}:
\begin{eqnarray}\label{eq:PipsihighQQCD}
    \Pi^{c,b}_{\mbox{QCD}}(Q^2) &\sim& -\frac{1}{4\pi^2}  \log \frac{Q^2}{\nu^2} -\frac{3 m_Q^2}{2\pi^2 Q^2} \nonumber\\
    &&-\frac{1}{12Q^4} \langle \frac{\alpha_s}{\pi} G^2\rangle + ... \nonumber\\
\end{eqnarray}
The holographic coefficient at order $(1/Q^4)$ contains a contribution proportional to $a_{f2}^2$, generated by two insertions of the dimension-two deformation, and a contribution linear in $a_{f4}$. In the heavy-quark sector, the former is associated with perturbative mass corrections proportional to $m_Q^4$. Motivated by this separation, we identify the term linear in $a_{f4}$ with the gluon-condensate contribution. We adopt the same prescription in the light sector, where $a_{f2}$ represents an effective dimension-two correction rather than the physical light-quark mass.
Separating contributions according to their origin in the near-boundary expansion, neglecting the contribution of the chiral condensate, which is absent in this model, we adopt the phenomenological identification
\begin{equation}
    \langle \frac{\alpha_s}{\pi} G^2\rangle = \pm\frac{16}{\pi^2} a_{f4}
\end{equation}
where the plus is for light quarks and the minus for heavy quarks. 
A final remark regarding the OPE is in order. 
To perform a complete matching of higher-order perturbative contributions, specifically, terms involving higher powers of the quark mass $m_q$ alongside logarithmic corrections \cite{Chetyrkin:1997qi}, one must modify the dilaton ansatz so that it explicitly incorporates these logarithmic terms in the small-$z$ limit.

The free parameters of the holographic model ($\kappa$, $a_1$, $a_2$, $c_1$, $c_2$, $s_1$, $s_2$) are fixed by means of a numerical optimization procedure. 
For this purpose, we introduce the following loss function, designed to minimize the deviation from the phenomenological targets:
\begin{eqnarray}\label{eq:loss}
\mathcal L&=&\sum_{n=1}^2 \left[ \left(\frac{m_n-m_n^\mathrm{tar}}{\sigma_{m_n}}\right)^2
+\left(\frac{f_n-f_n^\mathrm{tar}}{\sigma_{f_n}}\right)^2\right] \nonumber\\
&&+ \left(\frac{G_2-G_2^\mathrm{tar}}{\sigma_{G_2}}\right)^2\nonumber\\
&&+\lambda_\mathrm{shape}\langle \phi_V^2\rangle+\lambda_\mathrm{param} \langle a_i^2\rangle,\nonumber\\
\end{eqnarray}
%\lambda_\mathrm{smooth}\langle(\Delta^2\log\widetilde f_V)^2\rangle
where $m^\mathrm{tar}_1$, $m^\mathrm{tar}_2$, $f^\mathrm{tar}_1$, and $f^\mathrm{tar}_2$ are the input masses and decay constants of the ground state and first radial excitation, $\sigma_{m_n}$ and $\sigma_{f_n}$  the corresponding uncertainties, while 
$G_2^\mathrm{tar} = (0.012)$ GeV$^4$ and $\sigma_{G_2}=0.004$ GeV$^4$.
$\langle \phi_V^2\rangle$ is the mean of $\phi_V(z)^2$ in the range of $z$ used in the numerical calculation, \emph{i.e.} $10^{-4}\leqslant z\leqslant z_{max}$ with $z_{max}=14,10,6$ for $\rho,\psi,\Upsilon$ mesons, respectively (units in GeV$^{-1}$). 
$\lambda_\mathrm{shape}$ prevents $\phi_V(z)$ becoming too large or having divergences, $\lambda_\mathrm{param}$ prevents the optimized amplitudes $a_i$ from growing unnecessarily large.
We use $\lambda_\mathrm{shape}= \lambda_\mathrm{param}=10^{-5}$.
%\lambda_\mathrm{smooth}=1,
We find that the value of the gluon condensate is highly sensitive to the coefficients $c_1$ and $s_1$; indeed, even minor variations in these parameters induce significant shifts in the condensate. 
In contrast, the other observables computed below are stable.

The following Sections are dedicated to light mesons, charmonia, and bottomonia phenomenology.

\subsection{Light vector mesons}\label{sec:lightmesons}
To reconstruct the gauge kinetic function $f_\rho(z)$ for the $\rho$ meson sector, we employ the optimization procedure detailed in the previous section, utilizing the following input data: $m^\mathrm{tar}_{1}=775.26$ MeV, $m^\mathrm{tar}_{2}=1465$ MeV, $f^\mathrm{tar}_{1}=155$ MeV, and $f^\mathrm{tar}_{2}=131$ MeV. 
The effective uncertainties used in the run are $\sigma_m=(1,25)$ MeV and $\sigma_f=(2,10)$ MeV \cite{ParticleDataGroup:2026aaa}.
We assumed a larger error compared to the experimental one in the cases in which the experimental error alone is overly restrictive and too small to be realistically applied to this kind of theoretical models.
Masses and decay constants of radial excitations with $n>2$ are then predictions of the reconstructed one-channel vector tower.
Fitted parameters are: $\kappa = 0.226$ GeV$^2$, $a_1=-1.066$, $a_2=2.792$ (amplitudes), $c_1=0.717$ GeV$^{-1}$, $c_2=2.232$ GeV$^{-1}$ (basis centers), $s_1=0.487$ GeV$^{-1}$, $s_2=0.93$ GeV$^{-1}$ (basis widths), as shown in Table \ref{tab:params}.

\begin{table*}[tbp]
\centering
\caption{Parameters of the gauge kinetic function in Eq.~\eqref{eq:phiV_def} for $\rho$, $\psi$ and $\Upsilon$ mesons.}
\label{tab:params}
\begin{ruledtabular}    
\begin{tabular}{c c c c c c c c}  
\textbf{State} & \textbf{$\kappa$ [GeV$^2$]} & \textbf{$a_1$} & \textbf{$a_2$} & \textbf{$c_1$ [GeV$^{-1}$]} & \textbf{$c_2$ [GeV$^{-1}$]}& \textbf{$s_1$ [GeV$^{-1}$]} & \textbf{$s_2$ [GeV$^{-1}$]} \\
\noalign{\vspace{3pt}}
\colrule
\rule{0pt}{3ex}
 $\rho$ & 0.226 & -1.066 & 2.792 & 0.717 & 2.232 & 0.487 & 0.93 \\
 $\psi$ & 0.284 & 3.654 & -0.476 & 0.697  & 1.289 &  0.486 & 0.230 \\
 $\Upsilon$ & 0.3241 & 1.0262 & 1.7466 & 0.09937 & 0.3202 & 0.07536 & 0.3669 \\
\end{tabular}
\end{ruledtabular}
\end{table*}

\begin{figure}
    \centering
    \includegraphics[alt={Profile of gauge kinetic functions},width=0.9\linewidth]{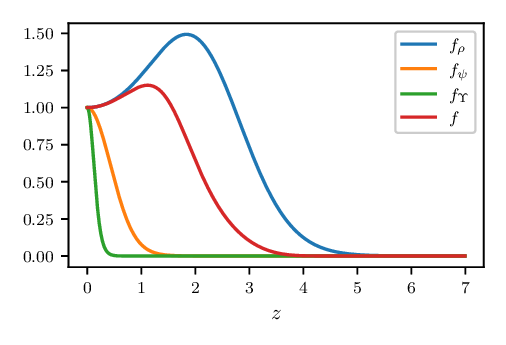}
    \caption{\label{fig:fV} Profile of the functions $f_\rho(z)$, $f_\psi(z)$, $f_\Upsilon(z)$, which have been reconstructed as illustrated in Section \ref{sec:newgaugefunc}, and $f(z)$ in Eq.~\eqref{fff}.}
\end{figure}

The resulting gauge kinetic function $f_V(z)=f_\rho(z)$ is positive definite and single-peaked. 
It is shown in Fig.~\ref{fig:fV}, compared with the function $f(z)$ in \eqref{fff}, determined in Ref.~\cite{Chen:2025goz}. 
They both increase at small $z$ and then decrease to zero, but $f_\rho(z)$ has a higher peak than the function $f(z)$ extracted from thermodynamic data. 

Table~\ref{tab:rho_spectrum_model} summarizes the masses and decay constants of the $\rho$-meson radial excitations obtained in this model, in comparison with available PDG values \cite{ParticleDataGroup:2026aaa}. 
The predicted decay constants decrease monotonically.
The final value of the gluon condensate is $G_2=0.0122$ GeV$^4$.
As expected, adopting the new $f_\rho(z)$ enhances the model's accuracy in replicating experimental masses.
The main systematic uncertainty is the one-channel nature of the vector equation: it cannot
resolve $S-D$ mixing, continuum effects, or a separate $D$-wave tower. 
In this setup, the $\rho(1700)$ is intentionally kept outside the fitted principal sequence, consistent with the interpretation of that state as mixed or non-principal in many spectroscopy analyses.

\begin{table*}[tbp]
\centering
\caption{Masses (Th mass) and decay constants (Th decay constant) of $\rho$-meson $n{}S$ radial excitations found in the model with reconstructed gauge kinetic function as described in Section \ref{sec:newgaugefunc}. 
Masses are compared to PDG values (Exp mass) \cite{ParticleDataGroup:2026aaa}. 
Units are MeV. The masses and decay constants of the states followed by a $^{(*)}$ were used as input in the  optimization algorithm.}
\label{tab:rho_spectrum_model}
\begin{ruledtabular}
\begin{tabular}{c c c c c}  
\textbf{State} & \textbf{Meson} & \textbf{Exp mass} & \textbf{Th mass}  & \textbf{Th decay constant} \\
%\noalign{\vspace{3pt}}
\colrule
\rule{0pt}{3ex}
$1S$ & $\rho(770)^{(*)}$ & $775.26 \pm 0.23$ &  775.3 & 155  \\
$2S$ & $\rho(1450)^{(*)}$ & $1465 \pm 25$ & 1467  & 131 \\
%\multirow{2}{*}{$3S$} 
%& $\rho(1700)$ & $1720 \pm 20$  & --- & --- \\
%\cline{2-5}  
$3S$ & $\rho(1900)$ & $1880 \pm 10$ (Ablikim 2022) & 1855 & 122 \\
$4S$ & $\rho(2150)$ & $2044 \pm 31\pm 4$ (Ablikim 2023) &  2141 & 117 \\
$5S$ & --- & --- & 2379 & 115 \\
\end{tabular}
\end{ruledtabular}
\end{table*}

The high-$Q^2$ expansion of the two-point function in Eq.~\eqref{eq:PihighQAdS} yields a non-vanishing, positive coefficient $a_{f2}$. 
In contrast, for massless quarks in standard QCD, this term is expected to vanish due to the absence of local, gauge-invariant operators of dimension two. 
Nevertheless, phenomenological fits to experimental data, such as $\tau$ lepton decays, often require the inclusion of an effective $1/Q^2$ contribution in the OPE of the vector current two-point function, a feature that has been extensively discussed in the literature \cite{Chetyrkin:1998yr,Narison:2009vy,Andreev:2006vy}.

The model can be employed to compute the leading-order HVP contribution to the anomalous magnetic moment of the muon ($g-2$).
In Ref.~\cite{Leutgeb:2022cvg}, it has been computed in different bottom-up holographic models, specifically in the hard-wall, the soft-wall model, and some generalizations.
The leading-order (LO) HVP contribution can be computed from \cite{Blum:2002ii,Leutgeb:2022cvg}:
\begin{equation}\label{eq:HVP}
    a_\mu^{\mbox{HVP,LO}} = 4 \alpha^2 \int_0^\infty \mathrm{d}K^2\, \hat f(K^2)\, (\Pi_{\mathrm{em}}(K^2)-\Pi_{\mathrm{em}}(0))\,,
\end{equation}
where
\begin{eqnarray}
    \hat f(K^2) &=& \frac{m_\mu^2 K^2 Z^3 (1-K^2 Z)}{1+m_\mu^2 K^2 Z^2} \nonumber\\
    Z &=& -\frac{K^2-\sqrt{K^4+4m_\mu^2 K^2}}{2 m_\mu^2 K^2}\,.
\end{eqnarray}
$m_\mu$ is the muon mass and $K^2$ the  squared Euclidean momentum.
The electromagnetic polarization is related to the correlator $\Pi(K^2)$ defined in \eqref{eq:twoptfunc} by
\begin{equation}
    \Pi^{u,d,s}_{\mathrm{em}}(K^2)=2\Tr[Q_{\mathrm{em}}^2]\, \Pi(K^2)
\end{equation}
where $Q_{\mathrm{em}}$ is the quark $u,d,s,$ charge matrix.
In \eqref{eq:HVP} the polarization has been renormalized by subtracting its value at $K^2=0$.
The  evaluation of $\Pi(K^2)$ in the $K^2 \to 0$ limit is numerically delicate due to the simultaneous vanishing of both the numerator and the denominator. 
Since the low-$K^2$ region provides a sizeable contribution to the HVP integral, an accurate determination of the polarization function in this regime is crucial. 
To circumvent this numerical instability, we extrapolated $\Pi(0)$ via a linear fit of $\Pi(K^2)$ computed at small $K^2$. 
This fit was also used to evaluate the integral in Eq.~\eqref{eq:HVP} in the extremely low-$K^2$ domain ($K^2 < 0.00012$ GeV$^2$).
To validate this procedure, we performed a small-$K^2$ expansion of the bulk-to-boundary propagator and solved the equation of motion at leading and next-to-leading order. 
This analytical approach yields the  same value for $\Pi(0)$.
Then, for $n_f=3$ quarks ($\Tr Q_{\mathrm{em}}^2 =2/3$) we find
\begin{equation}\label{eq:HVPuds}
    a_{\mu,uds}^{\mbox{HVP,LO}} = 7.051\times 10^{-8} \,.
\end{equation}
Notice that this prediction is slightly overestimated, since the strange-quark mass and strange mesons are a bit heavier than $u,d$ quarks and $\rho$ mesons, so they are expected to provide a lower contribution.
In the hard-wall and soft-wall models much lower values have been found for $a_{\mu,uds}^{\mbox{HVP,LO}}$ \cite{Leutgeb:2022cvg}.
In the 2025 White Paper on the anomalous magnetic moment of the muon \cite{Aliberti:2025beg}, the Standard Model prediction was updated by adopting a consolidated lattice-QCD evaluation for the leading-order HVP contribution, which caused a significant upward shift in the theoretical value and effectively resolved the long-standing tension with experimental measurements.
The  lattice-QCD determination of the total LO-HVP contribution adopted in Ref. \cite{Aliberti:2025beg} is $a_\mu^{\mbox{HVP,LO}} = 713.2(6.1) \times 10^{-10}$. This quantity includes connected and disconnected $u,d,s,c$ contributions and therefore is not directly equivalent to our flavor-symmetric connected $uds$ result. 
The connected lattice contributions from the $u,d$ and $s$ quarks are $659.5(4.7)\times10^{-10}$ and $53.37(11)\times10^{-10}$, respectively, giving a central-value sum of $712.87\times10^{-10}$. 
Our result is only a bit lower. However, because we employ the same vector correlator for all three light flavors, the model assigns $587.6\times10^{-10}$ to $u,d$ and $117.5\times10^{-10}$ to $s$. 
%The proximity of the summed results therefore involves a compensation between an underestimated \(u,d\) contribution and an overestimated strange contribution. 
Nevertheless, these outcomes show that a more refined holographic framework, capable of better capturing vector-meson phenomenology, yields a prediction for the HVP contribution that is in much better agreement with lattice QCD results.
A dedicated strange-sector gauge kinetic function would be required for a flavor-resolved comparison.

\begin{figure}[h!]
    \centering
    \includegraphics[width=0.9\linewidth]{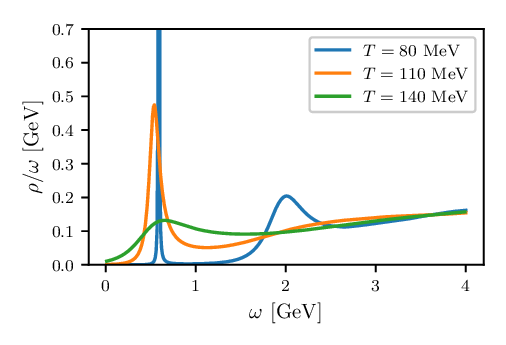}
    \caption{\label{fig:rhoSFmu0} Spectral function of the $\rho$ meson computed using the gauge function $f_\rho(z)$ for $\bar q=0$, $\mu=0$ and $T=80$ MeV (blue curve), $T=110$ MeV (orange curve), $T=140$ MeV (green curve).}
\end{figure}

\begin{figure}[h!]
    \centering
    \includegraphics[width=0.9\linewidth]{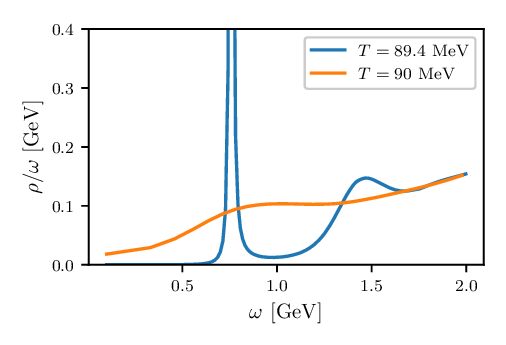}
    \caption{\label{fig:rhoSFmu1} Spectral function of the $\rho$ meson computed using the gauge function $f_\rho(z)$ for $\bar q=0$, $\mu=1$ GeV and $T=89.4$ MeV (blue curve), and $T=90$ MeV (orange curve).}
\end{figure}

At finite temperature and chemical potential, the $\rho$ meson spectrum is modified as previously shown. 
The $\rho$ spectral functions, computed from Eqs.~\eqref{eq:SF}-\eqref{eq:greenfun} replacing the function $f$ with $f_\rho$, are shown in Figs. \ref{fig:rhoSFmu0}-\ref{fig:rhoSFmu1} at $\bar q=0$ and  for $\mu=0$ and $\mu=1$ GeV, respectively, and for some values of temperature. 
The typical broadening of the peaks can again be observed. 
The $\rho$ meson melts at $T\sim 145$ MeV at $\mu=0$ and at $T\sim 90$ MeV at $\mu=1$ GeV, hence around the deconfinement phase transition.
To get this estimate, we have looked for the temperature at which the corresponding $\rho(\omega)$ becomes monotonic.
At large $\omega^2$, the behavior of the spectral function, that can be obtained from analytic continuation of Eq. \eqref{eq:PihighQAdS}, matches the one expected from free-quark Born approximation for vector mesons \cite{Petreczky:2005nh} (after adjusting for conventions):
\begin{equation}\label{eq:highq2rho}
    \rho(\omega) \sim \frac{N_c}{12\pi} \omega^2\,.
\end{equation}


\begin{figure}[h!]
    \centering
    \includegraphics[width=0.8\linewidth]{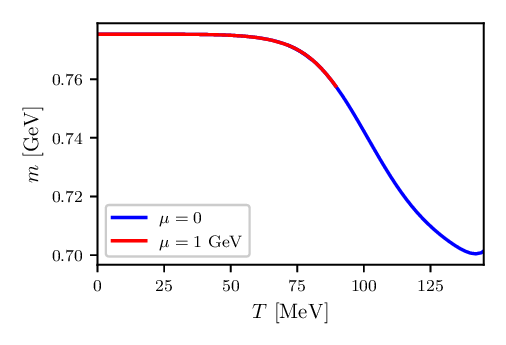}
    \caption{\label{fig:rhomassT} $\rho$ meson masses at finite temperature extracted from the spectral function through Eq.~\eqref{eq:BW} for $T>72$ MeV. 
    At smaller temperatures masses have been found by solving the corresponding Schr\"{o}dinger-like equation. }
\end{figure}

\begin{figure}[h!]
    \centering
    \includegraphics[width=0.8\linewidth]{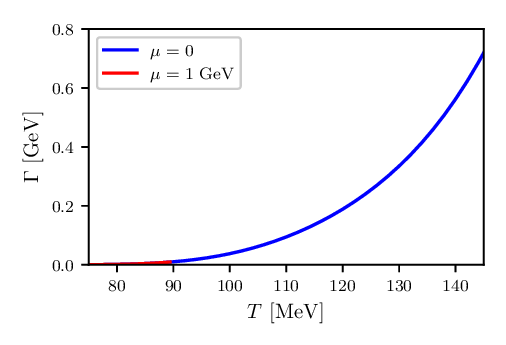}
    \caption{\label{fig:rhowidthT} $\rho$ meson widths at finite temperature extracted from the spectral function through Eq.~\eqref{eq:BW} for $T>72$ MeV. 
    At smaller temperatures widths are almost zero. }
\end{figure}

The peaks of the spectral functions can be fitted with a modified Breit-Wigner function as:
\begin{equation}\label{eq:BW}
    \rho_\mathrm{peak}(x) = \frac{\alpha_1 x^{\alpha_2} (1 + \alpha_3 x + \alpha_4 x^2)}{(x - M^2)^2 + M^2 \Gamma^2}
\end{equation}
with $x=\omega^2$, $\alpha_1$, $\alpha_2$, $\alpha_3$, $\alpha_4$, $M$, $\Gamma$ free parameters, $M$ and $\Gamma$ representing the mass and width of the resonance.  
The masses and widths of the $\rho$ meson thereby fitted at $\mu=0$ and $\mu=1$ GeV at varying temperature are shown in Figs. \ref{fig:rhomassT}-\ref{fig:rhowidthT}. 
At low temperatures, the width of the state is negligible, making the numerical computation of the spectral functions challenging. 
Furthermore, the position of the black-hole horizon, $z_h$, is sufficiently large that it does not affect the stability of the wavefunction. 
Consequently, we determined the low-temperature masses using the same procedure adopted at zero temperature, namely by solving the eigenvalue equation.
The $\rho$ mass decreases and the width increases in roughly the whole temperature range.


\begin{figure}[h!]
    \centering
    \includegraphics[width=0.8\linewidth]{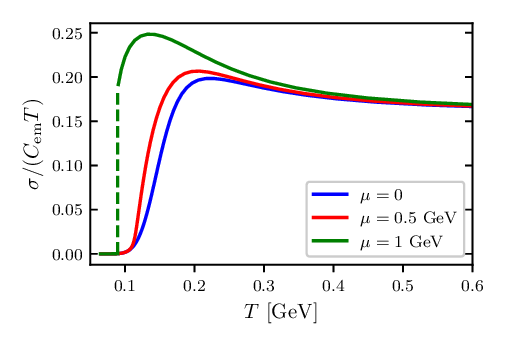}
    \caption{\label{fig:sigmanew} Electrical conductivity $\sigma/(C_{\mathrm{em}} T)$ for $\mu=0$ (blue curve), $\mu=0.5$ GeV (red curve), $\mu=1$ GeV (green curve), computed in the model with reconstructed gauge kinetic function $f_\rho(z)$.}
\end{figure}

The electrical conductivity computed with the new gauge kinetic function is shown in Fig.~\ref{fig:sigmanew}. 
A higher peak is found with respect to the conductivity computed from the thermodynamic gauge function (Fig.~\ref{fig:sigmatherm}), in line with the profiles in Fig.~\ref{fig:fV}.
At large temperature, when conformal symmetry is restored, we find $\sigma/(C_{\mathrm{em}} T)\to N_c/6\pi$, depending on the value we have fixed for $g_5^2$. 
In general, in holographic models, conformal invariance implies $\sigma \propto T$, with the constant determined by the normalization of the bulk gauge action.
D3/D7 models of Refs. \cite{Mateos:2007yp} and \cite{Mas:2008qs} give, for massless fundamental flavors at zero density, $\sigma/(C_{\mathrm{em}} T)\to N_c/4\pi$.
In contrast, the $\mathcal{N}=4$ SYM result scales as $N_c^2$, as indicated at the end of section \ref{sec:therm_mesons}, reflecting the $\mathcal{O}(N_c^2)$ adjoint degrees of freedom.
%A better matching with lattice data would be obtained by assuming a proper $f_Q(\phi)$ for the Maxwell-dilaton electric coupling, as done in \cite{Finazzo:2015xwa}.

\subsection{Charmonia}
To describe $c\bar c$ vector mesons, we introduce another vector field, dual to the $\bar c \gamma^\mu c$ current, and another gauge kinetic function $f_\psi(z)$ \cite{Zhang:2026zoz}. 
The action is given by Eq.~\eqref{eq:SV}, with $f_V$ replaced by $f_\psi$. 
To determine $f_\psi(z)$, we have again used the optimization algorithm described in the beginning of this Section. 
Input masses and decay constants of $J/\psi$ and $\psi(2S)$ are: $m_{J/\psi}=3097\pm 1$ MeV, $m_{\psi(2S)}=3686 \pm 1$ MeV, $f_{J/\psi}=415\pm 4$ MeV, $f_{\psi(2S)}=291\pm 9$ MeV \cite{ParticleDataGroup:2026aaa}.
For the masses we assumed a larger uncertainty compared to the experimental one.
The fitted parameters are indicated in Table \ref{tab:params}.
The resulting gauge kinetic function $f_\psi(z)$ is positive definite and monotonically decreasing. 
It is shown in Fig.~\ref{fig:fV} (orange curve).
The computed masses for the lowest-lying resonances in the $\psi$ spectrum are presented in Table \ref{tab:psimass}. 
These results are compared with the experimental masses of the observed states conventionally assigned to the $nS$ quantum numbers in potential models.
The fifth $S$-wave state has not been definitively identified experimentally.
Mixing among $S$-wave and $D$-wave states can occur.
The  value found for the gluon condensate is $G_2=0.0124$ GeV$^4$.

\begin{table}[tbp]
\centering
\caption{Masses (in MeV) of $\psi$-meson $n{}S$ radial excitations: comparison between PDG \cite{ParticleDataGroup:2026aaa} (Exp mass) and this work (Th mass). 
The masses of the states followed by a $^{(*)}$ were used as input in the  optimization algorithm.}
\label{tab:psimass}
\begin{ruledtabular}
\begin{tabular}{c c c c c}  
\textbf{State} & \textbf{Meson} & \textbf{Exp mass} & \textbf{Th mass} \\
\colrule
\rule{0pt}{3ex}
$1S$ & $J/\psi^{(*)}$ & $3096.900 \pm 0.006 $ &  3097 \\
$2S$ & $\psi(2S)^{(*)}$ & $ 3686.097 \pm 0.010$  & 3686 \\
$3S$ & $\psi(4040)$ & $ 4040\pm 4 $  & 4054 \\
$4S$ & $\psi(4415)$ & $ 4415 \pm 5 $ &  4320 \\
$5S$ & --- & --- &  4549 \\
\end{tabular}
\end{ruledtabular}
\end{table}

Decay constants $f_n$ are shown in Table \ref{tab:psiDC}. 
Their values can be used to compute the dilepton decay width from \cite{Giannuzzi:2008pv}
\begin{equation}\label{eq:gammall}
    \Gamma_{\ell^+\ell^-} = \frac{4\pi Q_c^2 \alpha^2 f_n^2}{3 m_n} \,,
\end{equation}
where $Q_c=2/3$ is the charge of the charm quark. 
The comparison between the values of $ \Gamma_{\ell^+\ell^-}$ in this model and the experimental ones is shown in Table \ref{tab:psiDC}. 
The decay constants and widths keep decreasing with the radial excitation number, however for highly radially excited charmonium states the predicted widths exceed the experimental values.
This discrepancy could be due to $S$–$D$ wave mixing, uncertain experimental state assignments, or the opening of the $D\bar{D}$ meson channels once the states cross the open-charm threshold.

\begin{table}[tbp]
\centering
\caption{Decay constants (in MeV) of $\psi$-meson $n{}S$ radial excitations. 
The two-lepton decay widths (in keV) computed from Eq.~\eqref{eq:gammall}, shown in the third column, are compared to experimental values extracted from \cite{ParticleDataGroup:2026aaa}, shown in the last column.}
\label{tab:psiDC}
\begin{ruledtabular}
\begin{tabular}{c c c c}  
\textbf{State} & \textbf{Decay constant} & \textbf{Th  $\Gamma_{\ell^+\ell^-}$} &  \textbf{Exp $\Gamma_{\ell^+\ell^-}$ } \\
\colrule
\rule{0pt}{3ex}
$1S$ &  415  & 5.52 & 5.53 \\
$2S$ &   291 & 2.28 & 2.27 \\
$3S$ &   234 & 1.35 & 0.857 \\
$4S$ &   225 & 1.16 & 0.352 \\
$5S$ &  221  & 1.06 & --- \\
\end{tabular}
\end{ruledtabular}
\end{table}

The two-point function of the $\bar c\gamma_\mu c$ current is given by:
\begin{equation}\label{eq:twoptfunccharm}
    \Pi(Q^2) = -\frac{2}{g_5^2 Q^2} \lim_{z\to 0} \frac{e^A}{z}\, f_\psi\, V\, \partial_z V\,,
\end{equation}
including a factor 2 with respect to the one in the light sector in Eq. \eqref{eq:twoptfuncNEW} due to the absence of the trace of Gell-Mann matrices.
By comparing the order $Q^{-2}$ of the high-$Q^2$ expansions of the two-point functions in Eqs. \eqref{eq:PihighQAdS} (multiplied by a factor 2), and \eqref{eq:PipsihighQQCD},  we find $m_c=0.95$ GeV, to be compared with $m_c(3 \mbox{ GeV}) =  0.986(13)$ GeV found in QCD sum rules \cite{Chetyrkin:2009fv}.   

The contribution of the charm quark to the HVP can be computed from Eq.~\eqref{eq:HVP} with
\begin{equation}\label{eq:emPIcb}
    \Pi^{c,b}_{\mathrm{em}}(K^2)=Q_{c,b}^2\, \Pi(K^2)\,,
\end{equation}
where $Q_c=2/3$ is the charm electric charge.
We find:
\begin{equation}
    a_{\mu,c}^{\mbox{HVP,LO}} = 1.39\times 10^{-9} \,. 
\end{equation}
As expected, the contribution from heavy quarks is severely suppressed. 
While the charm contribution is substantially smaller than that of the light quarks, it remains phenomenologically relevant. 
This is particularly true in light of the recent results from the Muon $g-2$ experiment at Fermilab, which achieved an unprecedented precision with an uncertainty of approximately $1.5 \times 10^{-10}$.
The SM prediction from lattice QCD is $a_\mu^{\mbox{HVP,LO}}(c) = 14.576(68) \times 10^{-10}$ \cite{Aliberti:2025beg}, so
%(roughly 0.2\% of the full HVP contribution), so charm quark's contribution must be taken into account and precisely determined.
our prediction is a bit lower than the SM result. 

\begin{figure}[h!]
    \centering
    \includegraphics[width=0.8\linewidth]{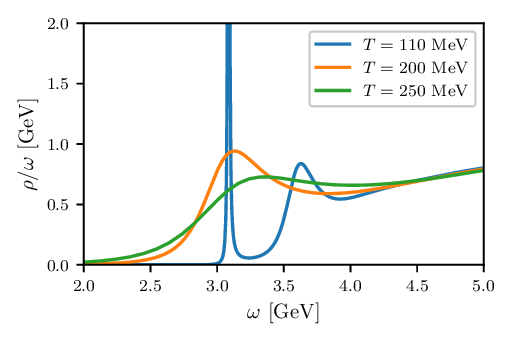}
    \caption{\label{fig:SFpsimu0} Spectral function of the $\psi$ mesons computed using the gauge function $f_\psi$, for $\bar q=0$, $\mu=0$ and $T=110$ MeV (blue curve), $T=200$ MeV (orange curve), $T=250$ MeV (green curve).}
\end{figure}

\begin{figure}[h!]
    \centering
    \includegraphics[width=0.8\linewidth]{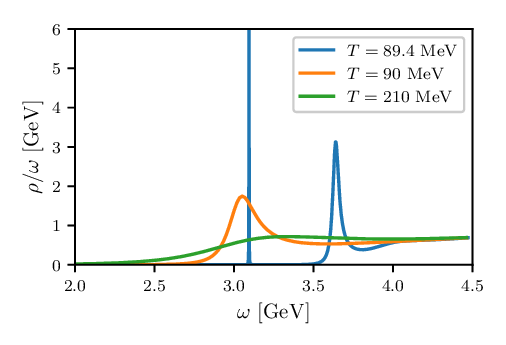}
    \caption{\label{fig:SFpsimu1} Spectral function of the $\psi$ mesons computed using the gauge function $f_\psi$, for $\bar q=0$, $\mu=1$ GeV and temperatures indicated in the legend.}
\end{figure}

Charmonium spectral functions at finite temperature and density have been computed in different holographic models, see \emph{e.g.} \cite{Braga:2017bml} and references therein.
The results obtained in this model are shown in Figs. \ref{fig:SFpsimu0}-\ref{fig:SFpsimu1} for $\mu=0$ and $\mu=1$ GeV, respectively.
The $J/\psi$ dissociates around $T\sim 250$ MeV at $\mu=0$ and $T\sim 210$ MeV at $\mu=1$ GeV, while $\psi(2S)$ dissociates soon after the phase transition in both cases.
Also in this case, the high-$\omega^2$ behavior in \eqref{eq:highq2rho} is very well reproduced, with a factor 2 due to the different normalization used for the heavy-quark vector current.


\begin{figure}[h!]
    \centering
    \includegraphics[width=0.8\linewidth]{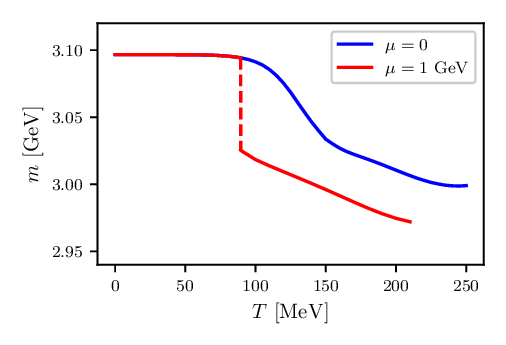}
    \caption{\label{fig:psimassesT} $J/\psi$ mass at varying temperature extracted from the spectral functions at $\bar q=0$.}
\end{figure}

\begin{figure}[h!]
    \centering
    \includegraphics[width=0.8\linewidth]{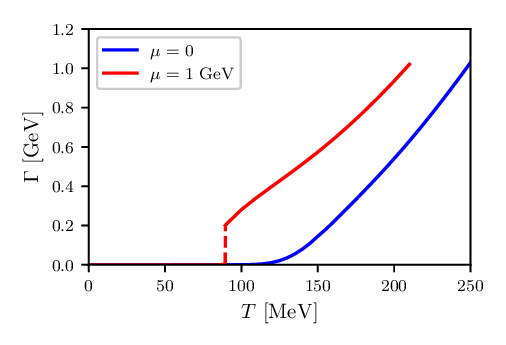}
    \caption{\label{fig:psiwidthsT} $J/\psi$ width at varying temperature extracted from the spectral functions at $\bar q=0$.}
\end{figure}

The masses and widths of the $J/\psi$ extracted from the spectral functions using Eq.~\eqref{eq:BW} are shown in Figs. \ref{fig:psimassesT}-\ref{fig:psiwidthsT} at varying temperature, for $\mu=0$ and $\mu=1$ GeV.
The masses at low temperatures ($T<90$ MeV) have been determined by solving the eigenvalue equation, as in the previous subsection.


\subsection{Bottomonia}
We have finally extended the framework to $\bar b b$ vector mesons.
In this case input data for determining the gauge function describing $\Upsilon$ mesons are: $m_{\Upsilon(1S)}=9460 \pm 1$ MeV, $m_{\Upsilon(2S)}=10023\pm 1$ MeV, $f_{\Upsilon(1S)}=702\pm 20$ MeV, $f_{\Upsilon(2S)}=497\pm 41$ MeV \cite{ParticleDataGroup:2026aaa}.
For the masses we assumed a larger uncertainty compared to the experimental one.
The obtained parameters are shown in Table \ref{tab:params}. 
The resulting gauge kinetic function $f_V(z)=f_\Upsilon(z)$ is positive definite and monotonically decreasing, much steeper than $f_\psi(z)$. 
It is shown in Fig.~\ref{fig:fV} (green curve).
Masses and decay constants obtained for the first five resonances are in Tables \ref{tab:upsilonmass} and \ref{tab:upsilonDC}, respectively, while the gluon condensate is $G_2\sim 0.0132$ GeV$^4$.

\begin{table}[tbp]
\centering
\caption{Masses (in MeV) of $\Upsilon$-meson $n{}S$ radial excitations: comparison between PDG (Exp mass) \cite{ParticleDataGroup:2026aaa} and this work (Th mass). 
The masses of the states followed by a $^{(*)}$ were used as input in the  optimization algorithm.}
\label{tab:upsilonmass}
\begin{ruledtabular}
\begin{tabular}{c c c c c}
\textbf{State} & \textbf{Meson} & \textbf{Exp mass} & \textbf{Th mass} \\
\colrule
\rule{0pt}{3ex}
$1S$ & $\Upsilon(1S)^{(*)}$ & $9460.40 \pm 0.10 $ &  9460 \\
$2S$ & $\Upsilon(2S)^{(*)}$ & $ 10023.4\pm 0.5$  & 10023 \\
$3S$ & $\Upsilon(3S)$ & $ 10355.1\pm 0.5 $  & 10314 \\
$4S$ & $\Upsilon(4S)$ & $ 10579.4 \pm 1.2 $ & 10531  \\
$5S$ & $\Upsilon(5S)$ & --- & 10721  \\
\end{tabular}
\end{ruledtabular}
\end{table}

\begin{table}[tbp]
\centering
\caption{Decay constants (in MeV) of $\Upsilon$-meson $n{}S$ radial excitations. 
The two-lepton decay widths  (in keV) computed from Eq.~\eqref{eq:gammall}, shown in the third column, are compared to experimental values extracted from PDG \cite{ParticleDataGroup:2026aaa}, shown in the last column.}
\label{tab:upsilonDC}
\begin{ruledtabular}
\begin{tabular}{c c c c}  
\textbf{State} & \textbf{Decay constant} & \textbf{Th $\Gamma_{\ell^+\ell^-}$} &  \textbf{Exp $\Gamma_{\ell^+\ell^-}$} \\
\colrule
\rule{0pt}{3ex}
$1S$ &  698  & 1.28  & 1.29 \\
$2S$ &  495  & 0.61 & 0.61 \\
$3S$ &  393  & 0.37 & 0.44 \\
$4S$ &  358  & 0.30 & 0.32 \\
$5S$ &  339  & 0.27 & --- \\
\end{tabular}
\end{ruledtabular}
\end{table}

From the OPE of the two-point function, as in the charm case, we find $m_b \sim 3.3$ GeV, to be compared with $m_b(10 \mbox{ GeV}) =  3.610(16)$ GeV obtained in QCD sum rules  \cite{Chetyrkin:2009fv}.

The contribution of the bottom quark to the HVP is roughly 20\% of the size of the current experimental error bar.
From Eqs. \eqref{eq:HVP} and \eqref{eq:emPIcb}, with $Q_b=-1/3$ the bottom electric charge and $f_\psi$ replaced by $f_\Upsilon$, we find:
\begin{equation}
    a_{\mu,b}^{\mbox{HVP,LO}} = 3.0\times 10^{-11} \,. 
\end{equation} 
The latter value matches the SM prediction from lattice QCD, which is $a_\mu^{\mbox{HVP,LO}}(b) = 0.30(02) \times 10^{-10}$ \cite{Aliberti:2025beg}. 

\begin{figure}[h!]
    \centering
    \includegraphics[width=0.8\linewidth]{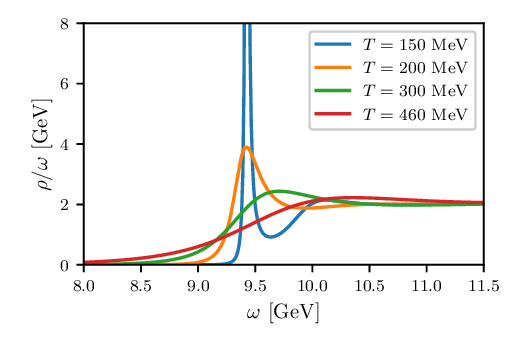}
    \caption{\label{fig:SFUpsilonmu0} Spectral function of the $\Upsilon$ mesons computed using the gauge function $f_\Upsilon$, for $\bar q=0$, $\mu=0$ and temperatures indicated in the legend.}
\end{figure}

\begin{figure}[h!]
    \centering
    \includegraphics[width=0.8\linewidth]{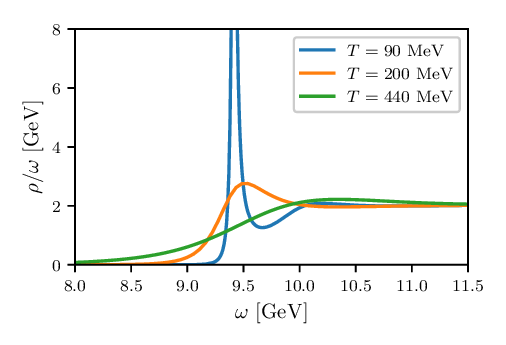}
    \caption{\label{fig:SFUpsilonmu1} Spectral function of the $\Upsilon$ mesons computed using the gauge function $f_\Upsilon$, for $\bar q=0$, $\mu=1$ GeV and temperatures indicated in the legend.}
\end{figure}

Bottomonium spectral functions at finite temperature and density are shown in Figs. \ref{fig:SFUpsilonmu0}-\ref{fig:SFUpsilonmu1} for $\mu=0$ and $\mu=1$ GeV, respectively.
The $\Upsilon(1S)$ dissociates around $T\sim 460$ MeV at $\mu=0$, corresponding to $\sim 3 \, T_c$, and $T\sim 440$ MeV at $\mu=1$ GeV, while $\Upsilon(2S)$ dissociates soon after the phase transition in both cases.
Also in this case, the high-$\omega^2$ behavior of the spectral function in \eqref{eq:highq2rho} is reproduced, with a factor 2 due to the different normalization used for the heavy-quark vector current.


\begin{figure}[h!]
    \centering
    \includegraphics[width=0.8\linewidth]{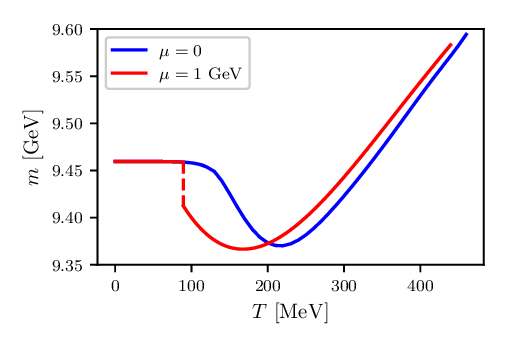}
    \caption{\label{fig:UpsilonmassesT} $\Upsilon$ masses at varying temperature extracted from the spectral functions at $\bar q=0$.}
\end{figure}

\begin{figure}[h!]
    \centering
    \includegraphics[width=0.8\linewidth]{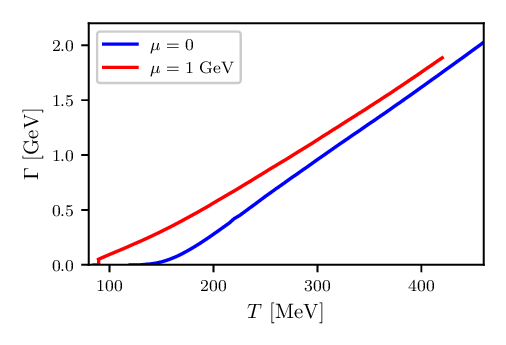}
    \caption{\label{fig:UpsilonwidthsT} $\Upsilon$ widths at varying temperature extracted from the spectral functions at $\bar q=0$.}
\end{figure}

The masses and widths of the $\Upsilon$ extracted from the spectral functions using Eq.~\eqref{eq:BW} are shown in Figs. \ref{fig:UpsilonmassesT}-\ref{fig:UpsilonwidthsT}.
The masses at low temperatures (\emph{i.e.} for $T<115$ MeV at $\mu=0$ and $T<89.5$ MeV at $\mu=1$ GeV) have been determined by solving the eigenvalue equation, as in the previous subsections.
Because bottomonium states exhibit higher dissociation temperatures in the QGP, they offer a wider temperature window to study in-medium modifications. 
In the high-temperature regime, our result exhibits an upward turn and eventually a positive thermal mass shift. A qualitatively similar non-monotonic trend, with the mass first decreasing and subsequently increasing, is visible in the red-square results of Fig. 1 of Ref.~\cite{Skullerud:2026sek}. However, the overall conclusion of that lattice study is a small negative thermal mass shift, and the magnitude of the effect is substantially smaller than in our model.


\section{Conclusions}
We have studied the $\rho$ meson spectrum in an AdS/QCD model both at zero and finite temperature and density.
The EMD model was fixed from thermodynamic data in \cite{Chen:2025goz}.
The predicted vacuum masses of the $\rho$ mesons systematically overestimate the experimental data. 
This discrepancy demonstrates that further theoretical refinements are needed to achieve a unified description of both thermodynamics and spectroscopy, at least under the assumption that the dilaton couples identically to the bulk fields dual to the baryon number and the vector fluctuations.

Subsequently, we determined the necessary behavior of the gauge kinetic function to successfully reproduce the masses and decay constants of the ground and first excited states. 
We observed that, as a function of the holographic coordinate $z$, this function must exhibit a significantly more pronounced peak.
The in-medium behavior of the $\rho$ meson was extracted from the spectral functions, which indicate that the resonance undergoes dissociation in the QGP at the deconfinement transition.

Furthermore, we extended our analysis to charmonia and bottomonia. 
To achieve this, we identified two distinct gauge kinetic functions capable of capturing the masses and decay constants of the lowest two states in each sector.

We computed the HVP contribution to the anomalous magnetic moment of the muon due to light and heavy quarks. 
Good agreement was found with lattice QCD calculations for charm and bottom contributions, and under the flavor-symmetric approximation in the light-quark sector.

In future work, we plan to extend this framework to the pseudoscalar and axial-vector meson sectors, enabling a detailed investigation of their spectra and the dynamics of chiral symmetry breaking. 
Furthermore, recent studies have shown that in certain holographic models the Melnikov-Vainshtein short-distance constraint on the hadronic light-by-light (HLbL) four-photon amplitude can be successfully fulfilled \cite{Mager:2025pvz}. 
However, the evaluation of the HLbL scattering tensor via  Green's functions in the standard soft-wall model is plagued by divergences \cite{Leutgeb:2025jmv}. 
It will therefore be  interesting to determine whether the refined model presented here can resolve these divergences and provide a consistent calculation of the HLbL contribution.

\section*{Acknowledgments}
This work is supported  by the National Natural Science Foundation of China (NSFC) Grant Nos: 12405154, and the European Union --- Next Generation EU through the research grant number P2022Z4P4B ``SOPHYA --- Sustainable Optimised PHYsics Algorithms: fundamental physics to build an advanced society'' under the program PRIN 2022 PNRR of the Italian Ministero dell'Universit\`a e Ricerca (MUR).
We thank L.~Cappiello, P.~Colangelo, F.~De~Fazio, J.~Mager and A.~Rebhan for fruitful discussions.
\bibliography{ref}

\end{document}